\documentclass[preprint,eqsecnum,prd,aps,nofootinbib]{revtex4}
\begin{document}

\def\cstok#1{\leavevmode\thinspace\hbox{\vrule\vtop{\vbox{\hrule\kern1pt
\hbox{\vphantom{\tt/}\thinspace{\tt#1}\thinspace}}
\kern1pt\hrule}\vrule}\thinspace}

\begin{center}
\bibliographystyle{article}
{\Large \textsc{Averaging procedure in variable-G cosmologies}}
\end{center}

\author{Vincenzo F. Cardone$^{1}$\thanks{
Electronic address: winnyenodrac@gmail.com},
Giampiero Esposito$^{2}$\thanks{
Electronic address: giampiero.esposito@na.infn.it}}

\affiliation{
${ }^{1}$I.N.A.F.,
Osservatorio Astrofisico di Catania, via S. Sofia 78, 95123 Catania, Italy \\
${ }^{2}$Istituto Nazionale di Fisica Nucleare, Sezione di
Napoli, Complesso Universitario di Monte S. Angelo, Via Cintia,
Edificio 6, 80126 Napoli, Italy}

\vspace{0.4cm}
\date{\today}

\begin{abstract}
Previous work in the literature had built a formalism for spatially
averaged equations for the scale factor, giving rise to an averaged
Raychaudhuri equation and averaged Hamiltonian constraint, which involve a
backreaction source term. The present paper extends these equations to
include models with variable Newton parameter and variable cosmological
term, motivated by the nonperturbative renormalization program for quantum
gravity based upon the Einstein\,-\,Hilbert action. We focus on the
Brans\,-\,Dicke form of the renormalization\,-\,group improved action
functional. The coupling between backreaction and spatially averaged
three\,-\,dimensional scalar curvature is found to survive, and a
variable\,-\,$G$ cosmic quintet is found to emerge. Interestingly, under
suitable assumptions, an approximate solution can be found where the early
universe tends to a FLRW model, while keeping track of the original
inhomogeneities through three effective fluids. The resulting
qualitative picture is that of a universe consisting of baryons only,
while inhomogeneities average out to give rise to the full dark-side
phenomenology. 
\end{abstract}

%\pacs{04.50.+h, 98.80.-k, 98.80.Es}

\maketitle
\bigskip
\vspace{2cm}

\section{Introduction}

Over the last ten years, the use of the effective average action
\cite{PHRVA-D57-971} and of the renormalization\,-\,group equations has
made it possible to obtain encouraging evidence in favour of Einstein's
gravity being renormalizable at non\,-\,perturbative level
\cite{PHRVA-D62-125021,PHRVA-D65-025013,CQGRD-19-483,PHRVA-D66-125001},
with all running couplings having a finite limit at large momenta. The
cornerstone of this program is the discovery of a new non\,-\,Gaussian
ultraviolet fixed point, besides the trivial one at the origin
\cite{HEP-TH-0511260,PHRVA-D73-041501,PRLTA-97-221301,07093851}. The
resulting picture seems to be as follows\,: at sub\,-\,Planckian distances,
spacetime is a fractal. It can be thought of as a self\,-\,similar
hierarchy of superimposed Riemannian manifolds of any curvature. As one
considers larger length scales where the renormalization\,-\,group running
of the gravitational parameters comes to a halt, the spacetime ripples
gradually disappear, and a classical four\,-\,dimensional spacetime
manifold is recovered \cite{HEP-TH-0511260}.

In the simplest possible terms, the {\it renormalization group improvement}
consists in the modified Einstein equations \cite{PHRVA-D65-043508}

\begin{equation}
R_{\mu\nu}-\frac{1}{2}g_{\mu\nu}R + \Lambda(k)g_{\mu\nu} = 8\pi G(k)
T_{\mu\nu},
\label{(1.1)}
\end{equation}
where the Newton parameter $G$ and cosmological term $\Lambda$ are now
dependent on the scale $k$, $k$ being the running cut\,-\,off of the
renormalization group equation \cite{PHRVA-D57-971}. In cosmology, the
dynamical evolution is determined by a set of renormalization group
equations by means of the cut\,-\,off identification $k = k(t)$ which
relates the energy scale of the running cutoff $k$ of the renormalization
group, with the cosmic time $t$. In \cite{PHLTA-B527-9} it has been shown
that, in a cosmological setting, the correct cutoff identification is $k
\propto t^{-1}$; it is thus possible to determine $G(k(t))$ and
$\Lambda(k(t))$ in Eq. (1.1) once a renormalization\,-\,group trajectory is
determined.

At a deeper level, one integrates out all fluctuations with momenta larger
than a cutoff ${\overline k}$, and one takes them into account by means of
a modified dynamics for remaining fluctuation modes with momenta smaller
than ${\overline k}$. This modified dynamics is generated by a
scale\,-\,dependent effective action $\Gamma_{k}$, whose $k$\,-\,dependence
is ruled by the exact renormalization\,-\,group equations. Flow equations
can be used for a complete quantization of fundamental theories. On
denoting by $S$ their classical action, one imposes the initial condition
$\Gamma(k = \kappa) = S$ at the scale of the ultraviolet cutoff $\kappa$,
and one exploits the renormalization\,-\,group equation to evaluate this
averaged effective action $\Gamma_{k}$ for all $k < \kappa$, and one then
takes the limits $k \rightarrow 0$ and $\kappa \rightarrow \infty$. In the
case of fundamental theories, the continuum limit $\kappa \to \infty$
exists after having renormalized finitely many parameters in the action,
and is evaluated at a non\,-\,Gaussian fixed point of the
renormalization\,-\,group flow. Such a new fixed point replaces the
Gaussian fixed point which, at least implicitly, underlies the construction
of theories which are instead perturbatively renormalizable
\cite{IMPAE-A20-2358}.

Notwithstanding its merits (see, however, the criticism expressed in
\cite{08053089}), the renormalization group approach has only been applied,
so far, to a strictly homogenous and isotropic universe. As is well known,
however, the matter distribution in the observable universe may be
considered homogenously distributed only on large scales so that it is
interesting to investigate how this can affect the
renormalization\,-\,group equations. Needless to say, lacking a detailed
description of the matter distribution, we can only consider a kind of {\it
average universe} obtained by averaging out the inhomogeneities.
Fortunately, over the last decade, progress has been made on the
longstanding problem of how to average a general inhomogeneous model. In
particular, the work in \cite{GRGVA-32-105} considers an irrotational fluid
motion with the associated Einstein equations (in $c = 1$ units)

\begin{equation}
R_{\mu \nu}-{1\over 2}g_{\mu \nu}R+\Lambda g_{\mu \nu} =8 \pi G \rho
u_{\mu} u_{\nu} .
\label{(1.2)}
\end{equation}
A flow\,-\,orthogonal coordinate system $x^{\mu}=(t,X^{k})$ is chosen, i.e.
Gaussian normal coordinates comoving with the fluid. On writing
$x^{\mu}=f^{\mu}(X^{k},t)$, one has $u^{\mu}={\partial f^{\mu}\over
\partial t}=(1,0,0,0)$ and $u_{\mu}={\partial f_{\mu}\over \partial
t}=(-1,0,0,0)$, where $t$ is proper time. The spacelike hypersurfaces of
constant $t$ are assumed to foliate the spacetime manifold, and their first
fundamental form, i.e. the spatial 3\,-\,metric $g_{ij}$, is used to define
\begin{equation}
J(t,X^{i}) \equiv \sqrt{{\rm det} \; g_{ij}} .
\label{(1.3)}
\end{equation}
The spatial averaging of a scalar field $\psi$ as a function of Lagrangian
coordinates and time on an arbitrary compact portion $D$ of the fluid is
defined by the volume integral (cf. \cite{PHRVA-D52-4393})
\begin{equation}
\langle \psi(t,X^{i}) \rangle_{D} \equiv
{1\over V_{D}} \int_{D} \psi(t,X^{i}) J \; d^{3}X ,
\label{(1.4)}
\end{equation}
where the volume $V_{D}$ is obtained as
\begin{equation}
V_{D}(t) \equiv \int_{D}J d^{3}X .
\label{(1.5)}
\end{equation}
A key role in the formalism is played by a dimensionless effective scale
factor
\begin{equation}
a_{D}(t) \equiv \left({V_{D}(t)\over V_{D}(t_{0})}\right)^{\frac{1}{3}} ,
\label{(1.6)}
\end{equation}
in terms of which the averaged expansion rate takes the form (the dot being
used for the total derivative with respect to $t$)
\begin{equation}
\langle \theta \rangle_{D} = {{\dot V}_{D}\over V_{D}} = 3{{\dot a}_{D}\over
a_{D}}.
\label{(1.7)}
\end{equation}
In \cite{GRGVA-32-105}, the scale factor $a_{D}$ obeys the averaged
Raychaudhuri equation
\begin{equation}
3{{\ddot a}_{D}\over a_{D}}+4 \pi G \langle \rho \rangle_{D} - \Lambda =
Q_{D} = {2\over 3} \langle (\theta - \langle \theta
\rangle_{D})^{2} \rangle_{D} - 2 \langle \sigma^{2}
\rangle_{D} ,
\label{(1.8)}
\end{equation}
where $Q_{D}$ is the backreaction source term and $\sigma^{2}$ is the rate
of shear, and the Hamiltonian constraint
\begin{equation}
3 \left({{\dot a}_{D}\over a_{D}}\right)^{2} - 8 \pi G \langle
\rho \rangle_{D} + {1\over 2} \langle { }^{(3)}R
\rangle_{D} - \Lambda = -{1\over 2}Q_{D},
\label{(1.9)}
\end{equation}
where ${ }^{(3)}R$ is the scalar curvature of the constant time spacelike
hypersurfaces used in the spacetime foliation.

Our aim here is to generalize the above averaging procedure to the
variable\,-\,$G$ cosmologies resulting from the renormalization\,-\,group
approach, to gain a better understanding of the conditions under which a
FLRW universe can be recovered from a renormalization\,-\,group approach
(rather than imposing a FLRW symmetry as has been done so far). For this
purpose, section 2 derives the Buchert average of the field equations
obtained from a Brans\,-\,Dicke approach to the renormalization\,-\,group
improved gravitational action. The integrability condition for the above
equations is obtained in section 3, while a variable\,-\,G cosmic quintet
is found to emerge in section 4. Section 5 studies accelerating patches and
stationary models, while a solution formula for the spatially averaged
scalar curvature is obtained in section 6. A particular solution of the
averaged equations is then investigated in Section 7, where we also
consider the case of a nearly homogenous universe through the useful
introduction of a set of effective fluids clarifying the role of the
different terms. Results and open problems are discussed in section 8. The
basic identities used for the evaluation of Buchert averages are described
in the appendix.

\section{The Buchert averaging method with variable $G$ and $\Lambda$
in a Brans\,-\,Dicke approach}

In light of equations (1.3)--(1.9), it is clear that the Buchert averaging
method aims at taking spatial averages of equations obtained from suitable
contractions and traces of tensor field equations. In particular, within
the framework of running\,-\,$G$ models, an interesting class is given by
the Brans\,-\,Dicke approach developed in \cite{PHRVA-D69-104022}. Here,
the Einstein\,-\,Hilbert action is renormalization\,-\,group improved by
replacing the Newton constant and the cosmological constant by scalar
functions $G$ and $\Lambda$ in the corresponding Lagrangian density
\cite{PHRVA-D69-104022}. The position dependence of $G$ and $\Lambda$ is
governed by a renormalization\,-\,group equation, and they have the status
of externally prescribed background fields (whereas in \cite{CQGRD-21-5005}
$G$ obeys an Euler\,-\,Lagrange equation), while the metric satisfies an
effective Einstein equation similar to that of Brans\,-\,Dicke theory
\cite{PHRVA-D69-104022}.

Following \cite{PHRVA-D69-104022}, we assume that the total action
functional $S$ of our theory consists of the Einstein\,-\,Hilbert part
$S_{EH}$ plus matter $S_{M}$ plus a term describing the four\,-\,momentum
carried by the fields $G(x)$ and $\Lambda(x)$, i.e. (hereafter $\varphi$
denotes the collection of matter fields coupled to gravity)
\begin{equation}
S = S_{EH}(g,G,\Lambda) + S_{M}(g,\varphi) + S_{\theta}(g,G,\Lambda).
\label{(2.1)}
\end{equation}
For the explicit form of the terms in (2.1) we refer the reader
to Ref. \cite{PHRVA-D69-104022}, for length reasons. The
resulting renormalization\,-\,group improved Einstein equation is given by
\begin{equation}
R_{\mu \nu}-{1\over 2}g_{\mu \nu}R=-\Lambda g_{\mu \nu}
+8\pi G T_{\mu \nu}+\tau_{\mu \nu}+\vartheta_{\mu \nu},
\label{(2.2)}
\end{equation}
where $\tau_{\mu \nu}$ results from the $x$ dependence of $G$
\cite{PHRVA-D69-104022}:
\begin{equation}
\tau_{\mu \nu}={1\over G^{2}}\left \{ 2(\nabla_{\mu}G)(\nabla_{\nu}G)
-G \nabla_{\mu} \nabla_{\nu}G -g_{\mu \nu}\Bigr[
2 (\nabla_{\rho}G)(\nabla^{\rho}G)
-G g^{\mu \nu}\nabla_{\mu}\nabla_{\nu}G \Bigr] \right \},
\label{(2.3)}
\end{equation}
while $\vartheta_{\mu \nu}$ is obtained from the functional derivative
of $S_{\theta}$ with respect to the metric, and can be taken to be of
the form \cite{PHRVA-D69-104022}
\begin{equation}
\vartheta_{\mu \nu}=-{3\over 2}{1\over G^{2}}\left[(\nabla_{\mu}G)
(\nabla_{\nu}G)-{1\over 2}g_{\mu \nu}(\nabla_{\rho}G)(\nabla^{\rho}G)
\right],
\label{(2.4)}
\end{equation}
which is a Brans\,-\,Dicke energy\,-\,momentum tensor for the {\it field}
$1/G$. For our purposes it is convenient to add explicitly the tensors
$\tau_{\mu \nu}$ and $\vartheta_{\mu \nu}$ to find
\begin{eqnarray}
\; & \; & \Phi_{\mu \nu} \equiv \tau_{\mu \nu}+\vartheta_{\mu \nu}
\nonumber \\
&=& {1\over G^{2}} \left \{ {1\over 2}(\nabla_{\mu}G)(\nabla_{\nu}G)
-G \nabla_{\mu} \nabla_{\nu} G +g_{\mu \nu}
\left[-{5\over 4}(\nabla_{\rho}G)(\nabla^{\rho}G)
+G \cstok{\ }G \right] \right \},
\label{(2.5)}
\end{eqnarray}
where $\cstok{\ }$ is the standard notation for the wave operator
$g^{\alpha \beta}\nabla_{\alpha} \nabla_{\beta}$.

The first equation we need on our way towards Buchert averages is the
Hamiltonian constraint or $G_{00}$ component, obtained from contraction of
the renormalization\,-\,group improved Einstein equation (2.2) with
$u^{\mu} u^{\nu}$, i.e.
\begin{equation}
u^{\mu} u^{\nu}\left(R_{\mu \nu}-{1\over 2}g_{\mu \nu}R
\right)={1\over 2} \left({ }^{(3)}R+K^{2}-K_{ij}K^{ij}\right)
= 8 \pi G \rho + \Lambda + u^{\mu}u^{\nu}\Phi_{\mu \nu},
\label{(2.6)}
\end{equation}
where we denote by $K_{ij}$ the extrinsic\,-\,curvature tensor of the
spacelike hypersurfaces that foliate the spacetime manifold and have taken
$T_{\mu \nu}$ of the form \cite{GRGVA-32-105}
\begin{equation}
T_{\mu \nu}=\rho u_{\mu} u_{\nu},
\label{(2.7)}
\end{equation}
while
\begin{eqnarray}
u^{\mu}u^{\nu}\Phi_{\mu \nu}&=& {1\over G^{2}}\left[{1\over 2}
G_{,0}^{2}-G G_{,00}+{5\over 4}(\nabla_{\rho}G)(\nabla^{\rho}G)
-G \cstok{\ }G \right] \nonumber \\
&=& -{3\over 4}\left({G_{,0}\over G}\right)^{2}
+\theta {G_{,0}\over G}+{5\over 4}g^{ij}{G_{,i}\over G}{G_{,j}\over G}
-{\bigtriangleup G \over G},
\label{(2.8)}
\end{eqnarray}
with $\bigtriangleup \equiv g^{ij}\nabla_{i}\nabla_{j}$ the
Laplacian operator (up to a sign). Hereafter, it is convenient to define
\begin{equation}
\psi_{G} \equiv \log \; G,
\label{(2.9)}
\end{equation}
and bear in mind that the spatial part of the spacetime four\,-\,metric is,
in our coordinates, the induced Riemannian three\,-\,metric $h_{ij} dx^{i}
\otimes dx^{j}$. The renormalization\,-\,group improved $00$ component of the
Einstein equation (2.2) reads therefore (we now write explicitly all
traces, since the desired Buchert averages are for traces of the Einstein
equations \cite{GRGVA-32-105})
\begin{eqnarray}
\; & \; & { }^{(3)}R+({\rm Tr}K)^{2}-{\rm Tr}K^{2} \nonumber \\
&=& 16 \pi G \rho +2\Lambda -{3\over 2}\psi_{G,0}^{2}
+\theta \psi_{G,0}+{5\over 4}
h({\rm grad}\psi_{G},{\rm grad}\psi_{G})
-{\bigtriangleup G \over G},
\label{(2.10)}
\end{eqnarray}
where, by definition,
\begin{equation}
({\rm grad}\psi_{G})_{i} \equiv {G_{,i}\over G}.
\label{(2.11)}
\end{equation}
Since the left\,-\,hand side of (2.10) has the same functional form as in
general relativity, the Buchert average (1.4) of (2.10) leads to (see
appendix)
\begin{eqnarray}
\; & \; & 3 \left({{\dot a}_{D}\over a_{D}}\right)^{2}
-8\pi \langle G \rho \rangle_{D}+{1\over 2}
\langle { }^{(3)}R \rangle_{D} -\langle \Lambda \rangle_{D}
\nonumber \\
&=& -{1\over 2}Q_{D}+ \langle \theta \psi_{G,0}\rangle_{D}
-{3\over 4} \langle \psi_{G,0}^{2} \rangle_{D}
+{5\over 4} \langle h({\rm grad}\psi_{G},{\rm grad}\psi_{G})
\rangle_{D}- \left \langle {\bigtriangleup G \over G}
\right \rangle_{D}.
\label{(2.12)}
\end{eqnarray}
The second equation we need can be obtained by contracting (2.2) with the
contravariant spacetime metric $g^{\mu \nu}$. Of course, this contains also
$R_{\; 0}^{0}$ already encoded, up to a sign, in (2.10), but the difference
between the full scalar curvature and $R_{\; 0}^{0}$ is the new equation we
need. We point out that, from (2.2) and (2.5), and exploiting the
Arnowitt\,-\,Deser\,-\,Misner identity with unit lapse function and
vanishing shift vector
\begin{equation}
{ }^{(4)}R={ }^{(3)}R+{\rm Tr}K^{2}+({\rm Tr}K)^{2}
-2{\partial \over \partial t}{\rm Tr}K \ ,
\label{(2.13)}
\end{equation}
one finds, by adding and subtracting ${\rm Tr}K^{2}$ in (2.13),
\begin{eqnarray}
{ }^{(4)}R &=& \left[{ }^{(3)}R+({\rm Tr}K)^{2}
-{\rm Tr}K^{2} \right]+2 {\rm Tr}K^{2}-2{\partial \over \partial t}
{\rm Tr}K \nonumber \\
&=& 8\pi G \rho +4 \Lambda -{9\over 2}\psi_{G,0}^{2}
+3 \theta \psi_{G,0}+3 {G_{,00}\over G} \nonumber \\
&+& {9\over 2} h({\rm grad}\psi_{G},{\rm grad}\psi_{G})
-3{\bigtriangleup G \over G},
\label{(2.14)}
\end{eqnarray}
where we can exploit (2.10) and then take the Buchert average to
obtain eventually (see appendix)
\begin{eqnarray}
\; & \; & {2\over 3}\left[\left \langle {\rm Tr}K^{2}
-{\partial \over \partial t}{\rm Tr}K \right \rangle_{D} \right]
\nonumber \\
&=& 2 \left({{\dot a}_{D}\over a_{D}}\right)^{2}
+{2\over 3}{\partial \over \partial t}\langle \theta \rangle_{D}
-{2\over 3}Q_{D} \nonumber \\
&=& -{8 \pi \over 3} \langle G \rho \rangle_{D}
+{2\over 3} \langle \Lambda \rangle_{D}
-\langle \psi_{G,0}^{2} \rangle_{D}
+{1\over 3}\langle \theta \psi_{G,0} \rangle_{D}
+\left \langle {G_{,00}\over G} \right \rangle_{D}
\nonumber \\
&+& {2\over 3} \langle h ({\rm grad}\psi_{G},{\rm grad}\psi_{G})
\rangle_{D}-{1\over 3} \left \langle
{\bigtriangleup G \over G} \right \rangle_{D}.
\label{(2.15)}
\end{eqnarray}
If the FLRW symmetry is imposed, our equations (2.10) and (2.14) are in
full agreement with (3.18a) and (3.18b) of \cite{PHRVA-D69-104022},
respectively, i.e. ($H$ being the Hubble parameter ${{\dot a}\over a}$, and
$\chi=1,0,-1$ the curvature parameter, while bearing in mind that our
pressure parameter vanishes)
$$
H^{2}+{\chi \over a^{2}}={1\over 3}\Lambda+{8 \pi \over 3}G \rho
+\psi_{G,0}H-{1\over 4}\psi_{G,0}^{2},
$$
$$
H^{2}+2{{\ddot a}\over a}+{\chi \over a^{2}}=\Lambda
-{5\over 4}\psi_{G,0}^{2}+{G_{,00}\over G}+2 \psi_{G,0}H.
$$
Otherwise they express corrections involving the spatial gradient of
$G$. Note that $8 \pi \langle G \rho \rangle_{D}$ can be eliminated
in (2.15) with the help of (2.12), but this step is inessential.

Last, but not least, we have to take the Buchert average of the
on\,-\,shell consistency equation derived in \cite{PHRVA-D69-104022}, i.e.
\begin{equation}
4 \pi \langle G T_{\nu}^{\; \nu} u^{\mu}\nabla_{\mu}G \rangle_{D}
=\langle u^{\mu}\nabla_{\mu} (G \Lambda) \rangle_{D},
\label{(2.16)}
\end{equation}
where $u^{\mu}$ is the same vector used in (2.6).

\section{Integrability condition}

The integrability condition for our coupled system (2.12), (2.15) and
(2.16) is obtained after re\,-\,expressing (2.12) and (2.15) in the form
\begin{equation}
3 \left({{\dot a}_{D}\over a_{D}}\right)^{2}
-8 \pi \langle G \rho \rangle_{D}
+{1\over 2}\langle { }^{(3)}R \rangle_{D}
-\langle \Lambda \rangle_{D}=-{1\over 2}Q_{D}
+\langle F_{1} \rangle_{D},
\label{(3.1)}
\end{equation}
\begin{equation}
3{{\ddot a}_{D}\over a_{D}}+4\pi \langle G \rho \rangle_{D}
-\langle \Lambda \rangle_{D}=Q_{D}+{3\over 2}\langle F_{2}\rangle_{D},
\label{(3.2)}
\end{equation}
where we have defined
\begin{equation}
F_{1} \equiv \theta \psi_{G,0}-{3\over 4}\psi_{G,0}^{2}
+{5\over 4}h({\rm grad}\psi_{G},{\rm grad}\psi_{G})
-{{\bigtriangleup G} \over G},
\label{(3.3)}
\end{equation}
\begin{equation}
F_{2} \equiv {1\over 3}\theta \psi_{G,0}-\psi_{G,0}^{2}
+{G_{,00}\over G}+{2\over 3}h({\rm grad}\psi_{G},{\rm grad}\psi_{G})
-{1\over 3}{{\bigtriangleup G}\over G}.
\label{(3.4)}
\end{equation}
Following \cite{CQGRD-22-L113}, we now take the time derivative
of (3.1) and then use again (3.1) and (3.2), i.e. (hereafter,
all partial time derivatives acting on Buchert averages coincide
with total time derivatives of such spatial averages)
\begin{eqnarray}
\; & \; & {\partial \over \partial t}3 \left({{\dot a}_{D}\over a_{D}}
\right)^{2}=2{{\dot a}_{D}\over a_{D}} \left(
3{{\ddot a}_{D}\over a_{D}}
-3 \left({{\dot a}_{D}\over a_{D}}\right)^{2}\right)
\nonumber \\
&=& 2{{\dot a}_{D}\over a_{D}}\biggr[-4\pi \langle G \rho \rangle_{D}
+\langle \Lambda \rangle_{D}+Q_{D}+{3\over 2}\langle F_{2} \rangle_{D}
\nonumber \\
&-& 8\pi \langle G \rho \rangle_{D}+{1\over 2}
\langle { }^{(3)}R \rangle_{D}-\langle \Lambda \rangle_{D}
+{1\over 2}Q_{D}-\langle F_{1} \rangle_{D}\biggr]
\nonumber \\
&=& 2{{\dot a}_{D}\over a_{D}}\left[-12 \pi \langle G \rho \rangle_{D}
+{3\over 2}Q_{D}+{1\over 2} \langle { }^{(3)}R \rangle_{D}
+\left \langle {3\over 2}F_{2}-F_{1} \right \rangle_{D}\right]
\nonumber \\
&=& 8\pi \left[\left \langle {\partial \over \partial t} G \rho
\right \rangle_{D}+\langle G \rho \theta \rangle_{D}
-3{{\dot a}_{D}\over a_{D}}\langle G \rho \rangle_{D}\right]
\nonumber \\
&+& {\partial \over \partial t}\langle \Lambda \rangle_{D}
-{1\over 2}{\partial \over \partial t}
\langle { }^{(3)}R \rangle_{D}-{1\over 2}{\partial Q_{D}\over \partial t}
+{\partial \over \partial t}\langle F_{1} \rangle_{D},
\label{(3.5)}
\end{eqnarray}
where use has been made of the Buchert identity \cite{GRGVA-32-105}
\begin{equation}
{\partial \over \partial t}\langle \psi \rangle_{D}
-\left \langle {\partial \psi \over \partial t}\right \rangle_{D}
=\langle \psi \theta \rangle_{D}-\langle \psi \rangle_{D}
\langle \theta \rangle_{D}
\label{(3.6)}
\end{equation}
with $\psi=G \rho$. Now both sides of (3.5) contain the term $-24 \pi
(\dot{a}_D/a_D) \langle G \rho \rangle_{D}$, leading eventually to the
desired integrability condition
\begin{eqnarray}
\; & \; & {\partial Q_{D}\over \partial t}+6{{\dot a}_{D}\over a_{D}}Q_{D}
+{\partial \over \partial t}\langle { }^{(3)}R \rangle_{D}
+2{{\dot a}_{D}\over a_{D}}\langle { }^{(3)}R \rangle_{D}
\nonumber \\
&=& a_{D}^{-6}\left[{\partial \over \partial t}(a_{D}^{6}Q_{D})
+a_{D}^{4}{\partial \over \partial t}\Bigr(a_{D}^{2}
\langle { }^{(3)}R \rangle_{D}\Bigr)\right]
\nonumber \\
&=& 16 \pi \left[\left \langle {\partial \over \partial t}(G \rho)
\right \rangle_{D}+\langle G \rho \theta \rangle_{D}\right]
+2{\partial \over \partial t} \langle \Lambda \rangle_{D}
\nonumber \\
&+& 2 \left[{\partial \over \partial t}\langle F_{1} \rangle_{D}
+{{\dot a}_{D}\over a_{D}}\langle (2F_{1}-3F_{2}) \rangle_{D}
\right],
\label{(3.7)}
\end{eqnarray}
where, from the definitions (3.3) and (3.4),
\begin{equation}
2F_{1}-3F_{2}=\theta \psi_{G,0}+{3\over 2}\psi_{G,0}^{2}
-3{G_{,00}\over G}+{1\over 2}h({\rm grad}\psi_{G},{\rm grad}\psi_{G})
-{{\bigtriangleup G}\over G}.
\label{(3.8)}
\end{equation}

\section{The variable\,-\,G cosmic quintet and cosmic equations}

Inspired by the work in \cite{CQGRD-22-L113}, it is convenient to introduce
the following dimensionless quantities:
\begin{equation}
\Omega_{m}^{D} \equiv {8 \pi \langle G \rho \rangle_{D}\over 3H_{D}^{2}},
\; \Omega_{\Lambda}^{D} \equiv {\langle \Lambda \rangle_{D}\over 3H_{D}^{2}},
\; \Omega_{R}^{D} \equiv -{\langle { }^{(3)}R \rangle_{D} \over
6 H_{D}^{2}}, \; \Omega_{Q}^{D} \equiv -{Q_{D}\over 6 H_{D}^{2}}, \;
\Omega_{G}^{D} \equiv {\langle F_{1} \rangle_{D}\over 3H_{D}^{2}},
\label{(4.1)}
\end{equation}
where the definition of $\Omega_{G}^{D}$ has been suggested by the averaged
Hamiltonian constraint (3.1), and hereafter $H_D \equiv \dot{a}_D/a_D$ is
the effective Hubble parameter. In a homogeneous and isotropic universe
with constant $G$ and $\Lambda$, $\Omega_{m}^{D}$ and
$\Omega_{\Lambda}^{D}$ reduce to the usual matter and cosmological constant
density parameters, $\Omega_{R}^{D}$ reduces to $\Omega_{k}$, while
$\Omega_{Q}^{D}$ and $\Omega_{G}^{D}$ vanish, so that the standard FLRW
cosmology is then recovered. Equations (4.1) represent therefore the
definition of the {\it cosmic quintet} of density parameters for
variable\,-\,G inhomogeneous models of the Brans\,-\,Dicke type. Such an
interpretation is also confirmed on noting that, by virtue of (3.1), one
finds
\begin{equation}
\Omega_{m}^{D} + \Omega_{\Lambda}^{D} + \Omega_{R}^{D} + \Omega_{Q}^{D}
+ \Omega_{G}^{D} = 1.
\label{(4.2)}
\end{equation}
Pursuing the analogy with FLRW
cosmology, it is instructive to recast (3.1), (3.2) and (3.7) into an
almost Friedmann form. To this aim, we look for an effective density
$\rho_{\rm eff}^{D}$ and effective pressure $p_{\rm eff}^{D}$ such that
(3.1) and (3.2) read as
\begin{equation}
3H_{D}^{2}=\langle \Lambda \rangle_{D}+8 \pi \langle G \rangle_{D}
\rho_{\rm eff}^{D},
\label{(4.3)}
\end{equation}
\begin{equation}
3{{\ddot a}_{D}\over a_{D}}=\langle \Lambda \rangle_{D}
-4 \pi \langle G \rangle_{D}\Bigr(\rho_{\rm eff}^{D}+3p_{\rm eff}^{D}\Bigr).
\label{(4.4)}
\end{equation}
By comparison with (3.1) and (3.2), we solve for $\rho_{\rm eff}^{D}$
and $p_{\rm eff}^{D}$ to find
\begin{equation}
\langle G \rangle_{D}\rho_{\rm eff}^{D}=\langle G \rho \rangle_{D}
-{1\over 16 \pi}\langle { }^{(3)}R \rangle_{D}
-{1\over 16 \pi}Q_{D}
+{\langle F_{1} \rangle_{D} \over 8 \pi},
\label{(4.5)}
\end{equation}
\begin{equation}
\langle G \rangle_{D} p_{\rm eff}^{D}=-{1\over 16 \pi}Q_{D}
+{1\over 48 \pi}\langle { }^{(3)}R \rangle_{D}
-{1\over 24 \pi} \langle (F_{1}+3F_{2})\rangle_{D}.
\label{(4.6)}
\end{equation}

\section{Accelerating patches and stationary models}

From equation (3.2), inspired by the work in \cite{CQGRD-22-L113}, we
see that the condition for an accelerating patch $D$ of the universe
reads as
\begin{equation}
Q_{D} > 4\pi \langle G \rho \rangle_{D}-\langle \Lambda \rangle_{D}
-{3\over 2} \langle F_{2} \rangle_{D}.
\label{(5.1)}
\end{equation}
The definition of the density parameters in (4.1) can be used to
re-express (5.1) in the form
\begin{equation}
-\Omega_{Q}^{D}> {1\over 4} \Omega_{m}^{D}-{1\over 2}\Omega_{\Lambda}^{D}
-{3\over 4}{\langle F_{2} \rangle_{D}\over \langle F_{1} \rangle_{D}}
\Omega_{G}^{D}.
\label{(5.2)}
\end{equation}
If the domain $D$ is taken to be as large as our observable universe
\cite{CQGRD-22-L113}, the Hamiltonian constraint (4.2), jointly with
(5.2), yields
\begin{equation}
{3\over 2}\Omega_{\Lambda}^{D}+\Omega_{R}^{D}> 1-{3\over 4}\Omega_{m}^{D}
-\left({3\over 4}{\langle F_{2} \rangle_{D}\over
\langle F_{1} \rangle_{D}}+1 \right)\Omega_{G}^{D}.
\label{(5.3)}
\end{equation}

\subsection{Stationary models}

In our approach, in order to evaluate the quantities occurring in
the equations, stationarity of the whole universe model is assumed. 
Within this framework, the manifold $\Sigma$ we refer to hereafter
must be a finite-volume compact manifold, so that the global average
makes sense. We find therefore the global stationarity 
conditions (cf \cite{CQGRD-22-L113})
\begin{equation}
Q_{\Sigma}=4\pi \langle G \rho \rangle_{\Sigma}
-\langle \Lambda \rangle_{\Sigma}-{3\over 2}
\langle F_{2} \rangle_{\Sigma},
\label{(5.4)}
\end{equation}
\begin{equation}
\langle { }^{(3)}R \rangle_{\Sigma}=12 \pi \langle G \rho \rangle_{\Sigma}
-6 H_{\Sigma}^{2}+3 \langle \Lambda \rangle_{\Sigma}
+\left \langle \left({3\over 2}F_{2}+2 F_{1}\right)\right \rangle_{\Sigma},
\label{(5.5)}
\end{equation}
\begin{equation}
H_{\Sigma}={{\dot a}_{\Sigma}\over a_{\Sigma}}={C\over a_{\Sigma}},
\label{(5.6)}
\end{equation}
where $Q_{\Sigma}$ and $\langle { }^{(3)}R \rangle_{\Sigma}$ are now the
global kinematical backreaction and the globally averaged three-dimensional
spatial curvature, respectively, while the constant $C$ can be obtained
from the Hamiltonian constraint evaluated at the initial time. Interestingly,
if we now insert the global stationarity conditions (5.4)--(5.6) into the
formulae (4.5) and (4.6) for effective density and effective pressure,
we find some remarkable cancellations (i.e. vanishing coefficients of
$\langle G \rho \rangle_{\Sigma}, \langle F_{1} \rangle_{\Sigma},
\langle F_{2} \rangle_{\Sigma}$), leading to
\begin{equation}
\langle G \rangle_{\Sigma}p_{\rm eff}^{D}={1\over 8\pi}
\Bigr(\langle \Lambda \rangle_{\Sigma}-H_{\Sigma}^{2}\Bigr),
\label{(5.7)}
\end{equation}
\begin{equation}
\langle G \rangle_{\Sigma}\rho_{\rm eff}^{D}={1\over 8\pi}
\Bigr(3 H_{\Sigma}^{2}- \langle \Lambda \rangle_{\Sigma}\Bigr).
\label{(5.8)}
\end{equation}
Hence we find the cosmic equation of state
\begin{equation}
\langle G \rangle_{\Sigma}\left(p_{\rm eff}^{D}+{1\over 3}
\rho_{\rm eff}^{D}\right)={1\over 12 \pi}\langle \Lambda \rangle_{\Sigma}.
\label{(5.9)}
\end{equation}

On taking the time derivative of $Q_{\Sigma}$ and
$\langle { }^{(3)}R \rangle_{\Sigma}$ in (5.4) and (5.5), and then using
the identity (3.6) with $\psi=G \rho, D=\Sigma$, and again (5.4)--(5.6),
we find
\begin{equation}
\left({\partial \over \partial t}+3{C \over a_{\Sigma}}\right)
\Bigr(Q_{\Sigma}+\langle \Lambda \rangle_{\Sigma}+{3\over 2}
\langle F_{2} \rangle_{\Sigma}\Bigr)=4 \pi
\left[\left \langle {\partial \over \partial t}(G \rho) \right
\rangle_{\Sigma}+\langle G \rho \theta \rangle_{\Sigma} \right],
\label{(5.10)}
\end{equation}
\begin{eqnarray}
\; & \; & {\partial \over \partial t} \langle { }^{(3)}R \rangle_{\Sigma}
+9{C \over a_{\Sigma}}Q_{\Sigma}
+6{\partial \over \partial t}H_{\Sigma}^{2}
=12 \pi \left[\left \langle {\partial \over \partial t}(G \rho)
\right \rangle_{\Sigma}
+\langle G \rho \theta \rangle_{\Sigma} \right] \nonumber \\
&-& 9{C \over a_{\Sigma}}\left(\langle \Lambda \rangle_{\Sigma}
+{3\over 2} \langle F_{2} \rangle_{\Sigma} \right)
+3{\partial \over \partial t}\langle \Lambda \rangle_{\Sigma}
+{\partial \over \partial t}\left \langle \left({3\over 2}F_{2}
+2 F_{1}\right)\right \rangle_{\Sigma}.
\label{(5.11)}
\end{eqnarray}

The complete integral of (5.10) is given by the complete
integral of the homogeneous equation \cite{CQGRD-22-L113}
$$
\left({\partial \over \partial t}+3{C \over a_{\Sigma}(t)}\right)
Q_{\Sigma}=0
$$
plus a particular integral of the full equation, so that we can
write (the initial value of the left-hand side below being denoted
by $J_{\Sigma}(t_{i})$)
\begin{eqnarray}
\; & \; & \left(Q_{\Sigma}+\langle \Lambda \rangle_{\Sigma}
+{3\over 2} \langle F_{2} \rangle_{\Sigma}\right)(t)
={J_{\Sigma}(t_{i})\over a_{\Sigma}^{3}(t)} \nonumber \\
&+& 4 \pi \int_{t_{i}}^{t} \gamma(t,t')
\left[\left \langle {\partial \over
\partial t'} (G \rho) \right \rangle_{\Sigma}(t') + \langle G \rho
\theta \rangle_{\Sigma}(t') \right]dt',
\label{(5.12)}
\end{eqnarray}
where $\gamma(t,t')$ is the Green function of the operator
${\partial \over \partial t}+3{C \over a_{\Sigma}(t)}$. Since we
are here assuming ${\dot a}_{\Sigma}=C$, we are actually dealing
with the first-order operator
${\partial \over \partial t}+{3C\over (Ct+a_{i})}$,
with $a_{i} \equiv a_{\Sigma}(t_{i})$. The desired Green function
solves the equation
\begin{equation}
\left({\partial \over \partial t}+{3C\over (Ct+a_{i})}
\right)\gamma(t,t')=0 \; \forall t \not = t',
\label{(5.13)}
\end{equation}
and suffers a jump at $t=t'$ given by \cite{Lanczos1961}
\begin{equation}
\lim_{t \to {t'}^{+}}\gamma(t,t')-\lim_{t \to
{t'}^{-}}\gamma(t,t') =1.
\label{(5.14)}
\end{equation}
It therefore reads as (hereafter $\Theta$ is the step function)
\begin{equation}
\gamma(t,t')=(Ct+a_{i})^{-3C}\Bigr[\Theta(t-t')h_{1}(t')
+\Theta(t'-t)h_{2}(t')\Bigr],
\label{(5.15)}
\end{equation}
where
\begin{equation}
h_{1}(t')=(Ct'+a_{i})^{3C}, \; h_{2}(t')=0,
\label{(5.16)}
\end{equation}
i.e. one finds, $\forall t \not = t'$,
\begin{equation}
\gamma(t,t')=\Theta(t-t'){\left({t'+a_{i}/C \over
t+a_{i}/C}\right)}^{3C}.
\label{(5.17)}
\end{equation}

\section{A solution formula for the spatially averaged scalar curvature}

Note now that, by virtue of (5.10), the equation (5.11) can be recast
in the form
\begin{equation}
{\partial \over \partial t}\left(\langle { }^{(3)}R
\rangle_{\Sigma}-3Q_{\Sigma} -6 \langle \Lambda \rangle_{\Sigma}
-\langle (6F_{2}+2F_{1})\rangle_{\Sigma}\right)
=-6{\partial \over \partial t}H_{\Sigma}^{2}
=12{C^{3}\over a_{\Sigma}^{3}},
\label{(6.1)}
\end{equation}
where a remarkable cancellation of coefficients for
${C \over a_{\Sigma}}Q_{\Sigma}$ has occurred. On denoting by $\chi$ an
integration constant, and bearing in mind that
$a_{\Sigma}(t)=Ct+a_{i}$, equation (6.1) is solved by
\begin{equation}
\langle { }^{(3)}R \rangle_{\Sigma}=3Q_{\Sigma}
+6 \langle \Lambda \rangle_{\Sigma}
+\langle (6F_{2}+2 F_{1})\rangle_{\Sigma}+\chi
+12C^{3}\int_{t_{i}}^{t}{\Theta(t-t')\over (Ct'+a_{i})^{3}}dt',
\label{(6.2)}
\end{equation}
where $Q_{\Sigma}$ is given by the solution formula (5.12). The coupling
between spatially averaged scalar curvature and backreaction is therefore
found to survive.

\section{A particular assumption on the averaged equations}

In order to work out a general solution of the cosmic equations, we should
know how matter is spatially distributed and assume a metric for the
inhomogenous universe to compute the averaged quantities. Moreover, an
ansatz should be made for the spatial variation of $G$,
possibly consistent with the renormalization-group
equations, so that $F_1$ and
$F_2$ can be evaluated. Since such an information
is lacking by virtue of severe technical difficulties
in deriving $G(x,t)$, we can only look for a
particular solution under some reasonable assumptions.

A first step along this road can be done by differentiating both sides of
Eq. (4.3) with respect to the cosmic time $t$, thus yelding
\begin{displaymath}
3 \frac{\ddot{a}_D}{a_D} = \langle \Lambda \rangle_D + 4 \pi \langle G
\rangle_D \left ( \frac{1}{H_D} 
\frac{\partial \rho^{D}_{{\rm eff}}}{\partial t}
+ 2 \rho^D_{{\rm eff}} \right ) + \frac{1}{2 H_D} \left ( \frac{\partial
\langle \Lambda \rangle_D}{\partial t} 
+ 8 \pi \rho^D_{{\rm eff}} \frac{\partial
\langle G \rangle_D}{\partial t} \right ) .
\end{displaymath}
In a FLRW universe, the effective fluid should be replaced by the standard
source (matter and radiation) term. In this case, the standard continuity
equation holds and the above relation identically reduces to the Raychaudhuri
equation. Let us assume that our effective fluid can still satisfy, to a 
first approximation, the standard continuity equation, i.e.
\begin{equation}
\frac{\partial \rho^D_{{\rm eff}}}{\partial t} + 3 H_D (\rho^D_{{\rm eff}}
+ p^D_{{\rm eff}}) = 0 .
\label{eq: frwcont}
\end{equation}
In classical general relativity, this equation holds by construction
of the effective equations. In our approach, consistency of the theory
seems to demand that Eq. (7.1) should hold, but for example we cannot
yet say what would one expect if (7.1) had source terms. To study
(tiny) departures from (7.1), one might try to assume analyticity
of $G(x,t),\Lambda(x,t)$ and hence set up a perturbative scheme
for the evaluation of all averaged equations. This is an important
topic that deserves attention in a separate paper.

On solving with respect to $(1/H_D) \partial \rho^D_{{\rm eff}}/\partial t$,
the above relation becomes
\begin{displaymath}
3 \frac{\ddot{a}_D}{a_D} = \langle \Lambda \rangle_D - 4 \pi \langle G
\rangle_D (\rho^D_{{\rm eff}} + 3 p^D_{{\rm eff}})
+ \frac{1}{2 H_D} \left ( \frac{\partial
\langle \Lambda \rangle_D}{\partial t} + 8 \pi
\rho^D_{{\rm eff}} \frac{\partial
\langle G \rangle_D}{\partial t} \right ) .
\end{displaymath}
By equating to Eq. (4.4), we get the remarkable relation
\begin{equation}
\frac{\partial
\langle \Lambda \rangle_D}{\partial t} + 8 \pi
\rho^D_{{\rm eff}} \frac{\partial
\langle G \rangle_D}{\partial t} = 0,
\label{eq: lgder}
\end{equation}
which, for a homogenous and isotropic universe, reduces to
\begin{displaymath}
\dot{\Lambda} + 8 \pi \rho \dot{G} = 0,
\end{displaymath}
which has already been obtained in the literature \cite{PHLTA-B527-9}.

It is worth wondering whether Eq. (\ref{eq: frwcont}) can lead to a
constraint also on the functions $F_1$ and $F_2$.
For this purpose, let us first
note that, by virtue of the definitions of the effective fluid energy
density and pressure, one has
\begin{displaymath}
\langle G \rangle_D \frac{\partial \rho^D_{{\rm eff}}}{\partial t} =
\frac{\partial \langle G \rho \rangle_D}{\partial t} - \frac{1}{16 \pi}
\frac{\partial Q_D}{\partial t} - \frac{1}{16 \pi}
\frac{\partial \langle ^{(3)}R \rangle_D}{\partial t} + \frac{1}{8 \pi}
\frac{\partial \langle F_1 \rangle_D}{\partial t}
- \rho^D_{{\rm eff}} \frac{\partial
\langle G \rangle_D}{\partial t} ,
\end{displaymath}
\begin{displaymath}
\langle G \rangle_D (\rho^D_{{\rm eff}} + p^D_{{\rm eff}})
= \langle G \rho \rangle_D
- \frac{1}{8 \pi} Q_D - \frac{1}{24 \pi} \langle ^{3}R \rangle_D
+ \frac{1}{8 \pi}
\langle F_1 \rangle_D - \frac{1}{24 \pi} \langle (F_1 + 3 F_2) \rangle_{D} .
\end{displaymath}
Multiplying by $\langle G \rangle_D$ and inserting these relations, the
continuity equation (\ref{eq: frwcont}) for the effective fluid becomes
\begin{eqnarray}
\left ( \frac{\partial Q_D}{\partial t} + 6 H_D Q_D \right ) +
\left ( \frac{\partial \langle ^{(3)}R \rangle_D}{\partial t} +
2 H_D \langle ^{(3)}R \rangle_D \right ) & = & 16 \pi
\left ( \frac{\partial \langle G \rho \rangle_D}{\partial t} +
3 H_D \langle G \rho \rangle_D \right ) \nonumber \\ ~ & + & 2
\left ( \frac{\partial \langle F_1 \rangle_D}{\partial t} +
3 H_D \langle F_1 \rangle_D \right ) \nonumber \\ ~ & - & 2 H_D \langle
(F_1 + 3 F_2) \rangle_D - 16 \pi \rho^D_{{\rm eff}} \frac{\partial \langle
G \rangle_D}{\partial t} . \nonumber
\end{eqnarray}
Comparing this relation with the integrability condition (3.7) and making
use of the Buchert identity (3.6) with $\psi = G \rho$ and $\langle
\theta \rangle_D = 3 H_D$, we finally get
\begin{displaymath}
\frac{\partial \langle \Lambda \rangle_D}{\partial t}
+ 8 \pi \rho^D_{{\rm eff}} {\partial \over \partial t}
\langle G \rangle_{D}=
H_{D} \Bigr( 3 \langle F_1 \rangle_D - 2 \langle (F_1 + 3 F_2)
\rangle_D - \langle
(2 F_1 - 3 F_2) \rangle_{D} \Bigr),
\end{displaymath}
so that, since the left-hand side vanishes because of Eq. (\ref{eq: lgder}),
we obtain
\begin{equation}
3 \langle F_1 \rangle_D - 2 \langle (F_1 + 3 F_2) \rangle_D - \langle (2
F_1 - 3 F_2) \rangle_D = - \langle (F_1 + 3 F_2) \rangle_D = 0,
\label{eq: f1f2constr}
\end{equation}
having made use of the linearity of the mean operator. Summarizing, we can
therefore conclude that the conservation of the effective fluid implies two
remarkable relations. The first one, Eq. (\ref{eq: lgder}) is the
counterpart for an inhomogenous model of the renormalization\,-\,group
condition on the $G$ and $\Lambda$ time-derivatives. On the contrary,
Eq. (\ref{eq: f1f2constr}) is an interesting new constraint closely relating
the spatial variation of the two auxiliary functions $F_1$ and $F_2$.
Indeed, whatever is the exact spatial variation of $G$, the condition
$\langle F_1 \rangle_D = -3 \langle F_2 \rangle_D$ must hold. It is worth
stressing, however, that this constraint only refers to the averaged
quantities, so that it is impossible to infer any
information on the spatial variation of $G$.

In order to gain further insight, it is interesting to consider
the case of a nearly homogenous and isotropic universe. As a first step, we
introduce {\it three auxiliary effective fluids} with energy densities
\begin{equation}
\left \{
\begin{array}{l}
\displaystyle{\rho^D_M} \equiv
\displaystyle{\frac{\langle G \rho \rangle_D}{\langle G \rangle_D}
- \frac{3}{8 \pi} \frac{\langle F_2 \rangle_D}{\langle G \rangle_D}}, \\
~ \\
\displaystyle{\rho^D_Q} \equiv
\displaystyle{- \frac{1}{16 \pi} \frac{Q_D}{\langle G \rangle_D}}, \\
~ \\
\displaystyle{\rho^D_R} \equiv
\displaystyle{- \frac{1}{16 \pi} \frac{\langle ^{(3)}R
\rangle_D}{\langle G \rangle_D}
+ \frac{1}{8 \pi} \frac{\langle (F_1 + 3 F_2) \rangle_D}{\langle G
\rangle_D}}, \\
\end{array}
\right .
\label{eq: rhonew}
\end{equation}
and pressure given as
\begin{equation}
\left \{
\begin{array}{l}
\displaystyle{p^D_M} \equiv 0, \\
~ \\
\displaystyle{p^D_Q} \equiv
\displaystyle{- \frac{1}{16 \pi} \frac{Q_D}{\langle G \rangle_D}}, \\
~ \\
\displaystyle{p^D_R} \equiv
\displaystyle{\frac{1}{48 \pi} \frac{\langle ^{(3)}R
\rangle_D}{\langle G \rangle_D}
- \frac{1}{24 \pi} \frac{\langle (F_1 + 3 F_2) \rangle_D}{\langle G
\rangle_D}}, \\
\end{array}
\right .
\label{eq: pnew}
\end{equation}
so that the equations of state are $(w^D_M, w^D_Q, w^D_R) =
(0, 1, -1/3)$. As can be straightforwardly checked, one has
\begin{equation}
\left \{
\begin{array}{l}
\rho^D_{{\rm eff}} = \rho^D_M + \rho^D_Q + \rho^{D}_{R}, \\
~ \\ p^D_{{\rm eff}} = p^D_M + p^D_Q + p^{D}_{R}, \\
\end{array}
\right .
\label{eq: rhosum}
\end{equation}
which shows that the effective fluid may be viewed as consisting of three
distinct components. The first one, $\rho^D_M$, has zero pressure and
reduces to the standard matter term when we turn off the spatial variation
of $G$ (so that $F_1 = F_2 = 0$) and replace averaged quantities with
standard ones in a FLRW universe. We can therefore think of it as the matter
term of our model although, since we do not know anything about $F_1$ and
$F_2$, we cannot rule out a priori that $(-3/8 \pi) (\langle F_2
\rangle_D/\langle G \rangle_D)$ overcomes
$\langle G \rho \rangle_D/\langle G \rangle_{D}$,
thus giving rise to a negative energy density. The second component, with
energy density $\rho^D_Q$, has the same
equation of state as the {\it stiff matter}.
However, should the backreaction term $Q_D$ be positive, its energy density
is negative thus leading to a negative pressure acting as a variable
cosmological constant (eventually driving cosmic speed up). Finally, the
term with energy density $\rho^D_R$ has a
negative equation of state, but we cannot say
whether it acts as a speeding up factor. Indeed, 
we do not know whether the curvature term
overcomes the other term giving a positive or negative energy density. Note
that, should we assume the conservation of the effective fluid, then
Eq. (\ref{eq: f1f2constr}) follows and $\rho_D^R$ is positive-definite.
However, to obtain results of general nature, we prefer
{\it not to assume here a priori the validity} of Eq. (\ref{eq: frwcont}).
In other words, at this stage, we are considering effective 
fluids which make it unnecessary to consider Eq. (7.1).

On denoting with $w^D_{{\rm eff}}
\equiv p^D_{{\rm eff}}/\rho^D_{{\rm eff}}$ the
effective fluid equation of state,
and inverting Eqs. (\ref{eq: rhosum}), we get
\begin{equation}
\left \{
\begin{array}{l}
\displaystyle{\rho^D_Q = \frac{\rho^D_M}{4}
\left [ -1 + (1 + 3 w^D_{{\rm eff}})
\frac{\rho^D_{{\rm eff}}}{\rho^D_M} \right ]}, \\
~ \\
\displaystyle{\rho^D_R = \frac{3 \rho^D_M}{4}
\left [ -1 + (1 - w^D_{{\rm eff}})
\frac{\rho^D_{{\rm eff}}}{\rho^D_M} \right ]},
\end{array}
\right .
\label{eq: rhodrhoq}
\end{equation}
which are fully general. Let us now assume that the universe is nearly
homogenous and isotropic as it is expected to be, for instance, in its
infancy when perturbations had still to grow. In this case, we can
write
\begin{equation}
G \rho - (3/8 \pi) F_2 = G_N \rho_M^{FLRW} [1 + \Delta_M(t, X^i)],
\label{eq: grhofrw}
\end{equation}
\begin{equation}
G = G_N [1 + \Delta_G(t, X^i)] ,
\label{eq: gnfrw}
\end{equation}
where $G_N$ is the Newtonian gravitational constant, $\rho_M^{FLRW}$ the
dust matter energy density in a FLRW universe, and $\Delta_M(t, X^i)$ and
$\Delta_G(t, X^i)$ two unknown functions accounting for the small
perturbations induced by the inhomogeneities. It is worth noting that
Eqs. (\ref{eq: grhofrw}) and (\ref{eq: gnfrw}) may still be formally written
even if deviations from the FLRW universe are severe, the only difference
being that $\Delta_M(t, X^i)$ and $\Delta_G(t, X^i)$ are no longer much
smaller than 1. On averaging Eqs. (\ref{eq: grhofrw}) and (\ref{eq: gnfrw}),
the energy density of the matter\,-\,like term reads as
\begin{equation}
\rho^D_M = \frac{\langle G \rho - (3/8 \pi) F_2 \rangle_D}{\langle G \rangle_D}
= \rho_M^{FLRW} {\times} \frac{1
+ \langle \Delta_M \rangle_D}{1 + \langle \Delta_G \rangle_D} ,
\label{eq: rhodmfrw}
\end{equation}
while Eqs. (4.5) and the average of (\ref{eq: gnfrw}) give
\begin{equation}
\rho^{D}_{{\rm eff}} = \frac{3 H_D^2 
- \langle \Lambda \rangle_D}{8 \pi G_N
(1 + \langle \Delta_G \rangle_D)} .
\label{eq: rhodefffrw}
\end{equation}
On inserting Eqs. (\ref{eq: rhodmfrw}) and (\ref{eq: rhodefffrw}) into
Eqs. (\ref{eq: rhodrhoq}), we get
\begin{equation}
\left \{
\begin{array}{l}
\displaystyle{\rho^D_Q = \frac{1}{4} \left ( \frac{1
+ \langle \Delta_M \rangle_D}
{1 + \langle \Delta_G \rangle_D} \right )\left [ -1 + \frac{1 + \langle
\Delta_G \rangle_D}{1 + \langle \Delta_M \rangle_D}
\frac{(3 H_D^2 - \langle \Lambda \rangle_{D})}
{(8 \pi G_N \rho_M^{FLRW})} (1 + 3 w^D_{{\rm eff}}) \right ] \rho_M^{FLRW}}, ~
\\ ~ \\ \displaystyle{\rho^D_R = \frac{3}{4} \left ( \frac{1
+ \langle \Delta_M \rangle_D}
{1 + \langle \Delta_G \rangle_D} \right ) \left [ -1 + \frac{1 + \langle
\Delta_G \rangle_D}{1 + \langle \Delta_M \rangle_D}
\frac{(3 H_D^2 - \langle \Lambda \rangle_{D})}
{(8 \pi G_N \rho_M^{FLRW})} (1 - w^D_{{\rm eff}}) \right ] \rho_M^{FLRW}}.
\end{array}
\right . 
\label{eq: rhodrhorfrw}
\end{equation}
We can now work out an interesting property of these two effective fluids
by assuming a phenomenological ansatz for the term $3 H_D^2 - \langle
\Lambda \rangle_D$ and for the effective fluid
equation of state $w^D_{{\rm eff}}$. It is
reasonable to assume that, for very large $z$,
\begin{displaymath}
\left \{
\begin{array}{ll}
\displaystyle{w^D_{{\rm eff}}(z) \simeq 0}, \\ ~ \\
\displaystyle{3 H_D^2(z) - \langle \Lambda \rangle_D(z)
\simeq 8 \pi G_N \rho_M^{FLRW}(z)}.
\end{array}
\right .
\end{displaymath}
Equation (\ref{eq: rhodrhorfrw}) shows that,
in this limit, both $\rho^D_Q$ and
$\rho^D_R$ vanish so that the dust matter FLRW case is recovered in the
early universe whatever is the exact shape of the correction functions
$\Delta_M$ and $\Delta_G$. In other words,
we find that the early universe {\it tends}
to a FLRW model, but keeps track of the original inhomogeneities
through the two effective fluids with density $\rho^D_Q$ and $\rho^D_R$.
Actually, depending on the functional expression adopted for $H_D$, it is
also possible that $\langle \Delta_M \rangle$ vanishes identically at all
redshifts. We stress that the constraint $\langle \Delta_M \rangle_D(z) =
0$ does not imply that the universe is homogenous, but only that the matter
inhomogeneities average out to zero. As is clear from Eqs. (\ref{eq:
rhodrhorfrw}), in such a case, the two fictitious fluids $\rho^D_Q$ and
$\rho^D_R$ still contribute to the cosmic dynamics. In particular, should
$H_D$ lead to cosmic acceleration, we can therefore argue that these two
fluids arising from inhomogeneities drive the cosmic speed up in a matter
dominated universe.

\section{Concluding remarks and open problems}

Our original equations (2.12), (2.15), (2.16), (3.7), (4.5),
(4.6), (5.2), (5.4), (5.5), (5.10), (5.12), (6.2) provide a simple
but nontrivial application of the Buchert method for spatial
averages \cite{GRGVA-32-105, CQGRD-22-L113} to
renormalization-group improved action functionals with variable
$G$ and $\Lambda$ of the Brans--Dicke type
\cite{PHRVA-D69-104022}. It now remains to be seen whether our
averaged equations agree with the qualitative picture in general
relativity \cite{CQGRD-22-L113, PRLTA-90-031101}, according to
which backreaction effects point to a global instability of the
standard cosmological model, with exact solutions and perturbative
results modeling this instability lying in the right sector to
account for dark energy from inhomogeneities \cite{GRGVA-40-467,
0803-1401} (for a critical view, see however the work in
\cite{CQGRD-23-235}).

The definition of the effective fluids in section 7 makes it
clear what is the role played by the different inhomogeneity
terms introduced by the averaging procedure. In the early
universe, only the $\rho^{D}_{M}$ term survives and reduces to the
standard dust matter, so that the usual FLRW case is recovered. It
is however intriguing to note that, in the late epochs, the
effective matter fluid energy density $\rho^{D}_{M}$ is increased with
respect to the standard one because of the $(-3/8\pi) \langle F_{2}
\rangle_D/\langle G \rangle_{D}$ term. As such, one could speculate
that this latter component mimics an effective dark matter
component, while $\langle G \rho \rangle_D/\langle G \rangle_D$
accounts for the baryons. On the other hand, the two additional
fluids $\rho^D_Q$ and $\rho^D_R$ can both provide a negative
pressure so that they can drive accelerated expansion. The resulting
qualitative picture is that of a universe consisting of baryons only,
while inhomogeneities average out to give rise to the full dark-side
phenomenology. Needless to say, such a qualitative picture
must be substantiated by a detailed comparison with the data on
both the background expansion and the growth of structure, and hence
we cannot draw any definite conclusion at the moment. An alternative
possibility, from this point of view, is to compare our effective
fluids' picture with the recent literature on the morphon field
\cite{CQGRD-23-817, CQGRD-23-6379} in order to work out fruitful
analogies.

We may be criticized for having used the renormalization-group approach
only as a motivation, without ever exploiting it for the explicit
evaluation of the Buchert averages involving $G$ and $\Lambda$. On the
other hand, as far as we know, no-one has succeeded so far in obtaining
$G(x,t)$ and $\Lambda(x,t)$ from the exact integration of the
renormalization-group equation, and previous attention had always focused
on FLRW models where, by symmetry, $G$ and $\Lambda$ can only depend 
on the time variable. Without an explicit knowledge of the desired
$G(x,t)$ and $\Lambda(x,t)$, we cannot yet provide examples of our
averaging procedure. However, as we have stressed after Eq. (7.1),
a promising way out might be obtained by studying a power-series
expansion of $G$ and $\Lambda$, which is independent of any 
fixed-point assumption. We hope to be able to return to this point
in a separate paper.

Another interesting issue is whether our averaged equations with
variable Newton parameter can be relevant for the theoretical
scheme proposed in
\cite{NJOPF-9-377,PRLTA-99-251101,07123984} as yet another
alternative to dark energy. It would be also quite important to
repeat our analysis with the help of the covariant technique
developed in \cite{GRGVA-24-1015}.

\acknowledgments
G. Esposito is grateful to the Dipartimento di Scienze
Fisiche of Federico II University, Naples, for hospitality and
support. We are grateful to Claudio Rubano for conversations
and encouragement, and to Thomas Buchert
and Martin Reuter for enlightening correspondence.

\appendix
\section{Basic identities for the Buchert averages}

Relying upon \cite{GRGVA-32-105}, we express the extrinsic-curvature
tensor $K_{ij}$ of the spacelike hypersurfaces foliating the spacetime
manifold in the form
\begin{equation}
K_{ij}=-\sigma_{ij}-{\theta \over 3}g_{ij},
\label{(A1)}
\end{equation}
where the shear tensor is symmetric and traceless. Hence one finds the
simple but fundamental identities (hereafter $\sigma^{2} \equiv
{1\over 2}\sigma_{ij} \sigma^{ij}$)
\begin{equation}
{\rm Tr}K=-\theta,
\label{(A2)}
\end{equation}
\begin{equation}
{\rm Tr}K^{2}=2 \sigma^{2}+{\theta^{2}\over 3},
\label{(A3)}
\end{equation}
\begin{equation}
{1\over 2}\Bigr(({\rm Tr}K)^{2}-{\rm Tr}K^{2})={\theta^{2}\over 3}
-\sigma^{2}.
\label{(A4)}
\end{equation}
When we evaluate the Buchert average of the Hamiltonian constraint
(2.10), we exploit the identity (1.7) and the definition (1.8) to write
\begin{equation}
{1\over 3}\langle \theta^{2} \rangle_{D}-\langle \sigma^{2}
\rangle_{D}={Q_{D}\over 2}+{1\over 3}\langle \theta \rangle_{D}^{2}
={Q_{D}\over 2}+3 \left({{\dot a}_{D}\over a_{D}}\right)^{2}.
\label{(A5)}
\end{equation}

Along the same lines, the first equality in (2.15) is obtained by
virtue of
\begin{equation}
\left \langle {\rm Tr}K^{2}-{\partial \over \partial t}
{\rm Tr}K \right \rangle_{D}=2 \langle \sigma^{2} \rangle_{D}
+{1\over 3}\left[\langle \theta^{2} \rangle_{D}
-\langle \theta \rangle_{D}^{2}\right]
+{1\over 3}\langle \theta \rangle_{D}^{2}
+\left \langle {\partial \theta \over \partial t} \right
\rangle_{D},
\label{(A6)}
\end{equation}
where \cite{GRGVA-32-105}
\begin{equation}
\left \langle {\partial \theta \over \partial t} \right \rangle_{D}
={\partial \over \partial t} \langle \theta \rangle_{D}
-\langle \theta^{2} \rangle_{D}
+\langle \theta \rangle_{D}^{2}.
\label{(A7)}
\end{equation}
The identities (1.7), (A6) and (A7) lead eventually to the first
equality in (2.15).

\end{document}